# Comment on "Ultrathin Interface Regime of Core-Shell Magnetic Nanoparticles for Effective Magnetism Tailoring"

*Carlos Martinez-Boubeta\**

Freelancer in Bilbao, Gasaren Plaza 48007, Spain.

**ABSTRACT** Here we demonstrate that the main findings of the recent Nano Letter paper by Moon *et al.* (DOI: 10.1021/acs.nanolett.6b04016) are incorrect and unphysical, that is, there is no possibility of heating efficiencies about tens of kilowatts per gram ferrofluid. We also note that the key concept of their paper, a large enhancement of magnetic coercivity in hard-soft core-shell nanoparticles when the shell thickness is only a few atomic layers thick, mimics results from previous work in exchange biased thin films. And so the introduction of the *"enhanced spin canting phenomenon"* will prove to be superfluous.

# Introduction

There has been renewed interest in the use of magnetic nanoparticles to convert high frequency electromagnetic energy into heat. During the last decade, numerous examples in the field of catalysis,[1] aeronautical engineering,[2] and biomedical applications have flourished. In a recent Letter "Ultrathin Interface Regime of Core–Shell Magnetic Nanoparticles for Effective Magnetism Tailoring", Moon *et al.*[3] claim to have observed a new interface spin canting phenomenon in core-shell ferrrites that pushes heating efficiencies above 10,000 W/g. If this were true, it represents new opportunities for further tailoring nanoscale agents to be used in drug delivery and selective destruction of tumours by hyperthermia. But here we argue that these claims are utterly fallacious, and inconsistent with the facts of magnetism, and tend to confound promising medical technology and quackery.

Fanciful remedials with magnets are not new. Back in the early 1770s, Franz Anton Mesmer alleged healing usage of magnetic power by persuading a patient to drink a preparation containing iron.[4] Many trace magnetic therapy back to Paracelsus, who, on the other hand, was apparently well aware of the subconscious suggestion to the patient. Fooling the mind, at times, can be challenging. As an example we mention that throughout the story of the knight *Don Quixote*, a classic of Spanish literature, Sancho is promised an island for all of his service. A promise that held unrealistic expectations. Likewise, many research articles "promise" more than nanotechnology can deliver.[5,6] In particular, biomedical research, and especially when it comes to cancer, suffers from systemic flaws that fuel the debate of replicability and reproducibility.[7,8]

In this Comment we show first that heating efficiencies estimated from the temperature-versus-time curves are significantly different from the reported values by Moon *et al*. Second, we discuss the inconsistency between the magnetic energy product and evaluation of its potential for hyperthermia. Finally, we demonstrate that the most interesting conclusion of the paper —that a new interface phenomenon operates in the few atomic layer thick shell— is no longer supported.

## Results

**Cycling a magnetic field at certain selected frequencies and field amplitude can control the flow of heat from a particle**

As a first result we demonstrate in Figure 1 that the heating performances are two orders of magnitude lower than those estimated in ref. [3]. Efficiency is usually assessed by determining the specific loss power (SLP) as

$$SLP = \frac{CV}{m}\frac{dT}{dt} \quad (1)$$

with C, the medium specific heat capacity, m/V the ferrofluid concentration, and dT/dt the initial rate of change in temperature measured from the heating profile.

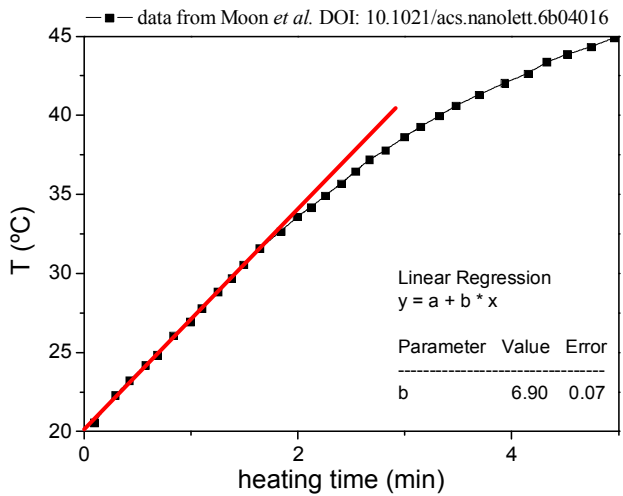

**Figure 1.** Experimental heating profile reported for the colloidal suspensions of 2 mg/ml cube-shaped nanoparticles with a $CoFe_2O_4$ core about 30 nm in size and 0.5 nm $MnFe_2O_4$ shell, at a field frequency of 30 kA/m at 500 kHz. Red line: initial linear trend (correlation coefficient r = 0.999).

In the paper by Moon *et al.* the sample referred to as CF@MF0.09 displays the highest dT/dt about 7 K per minute, for a concentration of 4 mg (nanoparticles) dispersed in 2 ml (toluene) tested in a field of 30 kA/m at 500 kHz (Fig. 4b of ref. [3]). The first thing that calls to our attention is the poor heating capability of these samples, given the fact that the declared values of power density would blow up the colloid. For comparison, concentration-normalized heating curves of iron oxide aqueous suspensions can be found in Roig *et al.*[9] Note in there that a sample showing 115 watts per iron gram attains similar temperature elevations to those reported in the Letter by Moon *et al.*, which is quite surprising given the several orders of magnitude difference in SLP values. Confirming our fears comes SLP ≈ 90 W/g, a generous estimate of what might be accomplished from CF@MF0.09 particles by using equation (1) and the saturated liquid heat capacity of toluene about 1500 kJ/m$^3$K [10]. In fact, our results are fully consistent with theoretical estimates by Fortin *et al.*[11] For instance, heating efficiency in a suspension of cobalt ferrite nanoparticles with sizes about 30 nm (which may compare to the core discussed here) is around 100 rather than 10,000 W/g.

A second point that deserves further attention relates to the maximum energy product, (BH)$_{max}$. It is an energy density with units of J/m$^3$ in the Systeme International, and it is equivalent to the area of the largest rectangle that can be inscribed under the demagnetization curve. Cycling the magnetization of a particle will correspondingly cause work which is dissipated in the environment as thermal energy. Therefore, (BH)$_{max}$ relates to the SLP as times the experimental frequency divided by the magnet density. Assuming a ferrite density about 5 g/cm$^3$ (as used in Lee *et al.*[12]), the energy product of 1,791 J/m$^3$ acquired for the CF@MF0.09 core-shell nanoparticles would make a maximum output of 180 W/g at 500 kHz.

**Of the strange manner in which the doubtful question of the "enhanced spin canting" is finally settled**

It is to be noted that the enhancement of the energy product in core/shell nanocomposites is no novelty in magnetism engineering. It was previously found that by tuning the dimensions of the FePt (hard) and $Fe_3O_4$ (soft) phases, $(BH)_{max}$ can get higher than the isotropic single-phase FePt counterpart by just adding thin soft shells (below 2 nm thick).[13] And with that in mind, allow us for a moment to deviate attention to thin film systems. Dusting the interface is one possibility to modify the exchange interactions between hard and soft materials. For instance, Ali *et al.*[14] reported the effects of inserting impurity layers of various elements into a Co/IrMn exchange biased bilayer. They observed an increase (decrease) of coupling for most of the magnetic (nonmagnetic) dusting layers, accompanied by an enhancement in $H_C$, with a broad peak below 5 Å of impurity that compares well with the case of sample CF@MF0.09 discussed above. Another interesting remark is that the percentage increase in Ali *et al.* was roughly proportional to the magnetization of the dusting element, as is the case here as well: the $H_C$ values of CF@$XFe_2O_4$ (X = Mn, Fe, Co, and Ni) for a shell volume fraction of 0.09, are 7.2, 6.9, 5.8, and 4.9 kOe, respectively. These are all higher than that of the CF core (3.3 kOe) and are said to be consistent with the number of unpaired d-electrons of shell atoms (see Fig. 3b in ref. [3]). In another example, Choi *et al.*[15] found that the maximum energy product of Sm–Co bilayers can be improved by 50% by adding an intermixed CoFe layer. And so it is clear that in exchange coupled hard–soft thin film systems the magnetization switching behavior and thus, the hysteresis loops, depend strongly on the dimensions of the soft phase (see López-Ortega *et al.*[16] for a review). Of particular help is the link established between layers thicknesses and the nucleation field

$$H_N = 2 \frac{f_{hard}K_{hard} + f_{soft}K_{soft}}{f_{hard}M_{hard} + f_{soft}M_{soft}} \quad (2)$$

where $f_{soft}$ corresponds to the shell volume fraction ($f_{shell} = V_{shell}/V_{total}$) in the paper by Moon *et al.* and $f_{hard} = 1 - f_{soft}$ is the core fraction. Same applies to the magnetic anisotropy (K) and saturation magnetization (M). Nucleation is the beginning of the magnetic reversal, which sets a lower limit for the coercivity, characterized by the average magnetic properties of the two layers. Therefore it is reasonable to assume that for a composite material, the effective characteristics are given by $K_{eff} = f_{soft}K_{soft} + (1-f_{soft})K_{hard}$ and $M_{eff} = f_{soft}M_{soft} + (1-f_{soft})M_{hard}$. We recall equation (2) is valid for a bilayer film system. Applying these concepts to the Stoner-Wohlfarth description, a random assemblage of core(hard)/shell(soft) nanoparticles will show a coercive field ($H_C$) about $K_{eff}/M_{eff}$ (several proportionality factors would apply for particles whose easy axes are aligned with the applied field, or whether the magnetization switching is controlled by, for instance, a cubic magnetocrystalline anisotropy). Still, it is worth noting that Moon *et al.* observed *"where the shell thickness is similar to the size of a crystalline unit cell, the population of canted spins increases"*. In this regard, we have previously shown that the inevitable presence of atomic steps in epitaxial systems implies an intrinsic morphological roughness at the interface between layers.[17] Given the unit cell size about 8.4 Å for the Mn-ferrite case in their paper, it is doubtful that shell thicknesses below 1 nm constitute a continuous layer and therefore, only a small fraction of the interfacial moments would contribute to bias. Thus, the cooperative coupling between the core and the incomplete shell should be phenomenologically renormalized by the nominal thickness of the shell in fractions of a monolayer.[18] We plot the simulated coercive field versus shell width in Figure 2. One can see that the agreement between predicted and measured values (Figure 2b from ref. [3]) is very good.

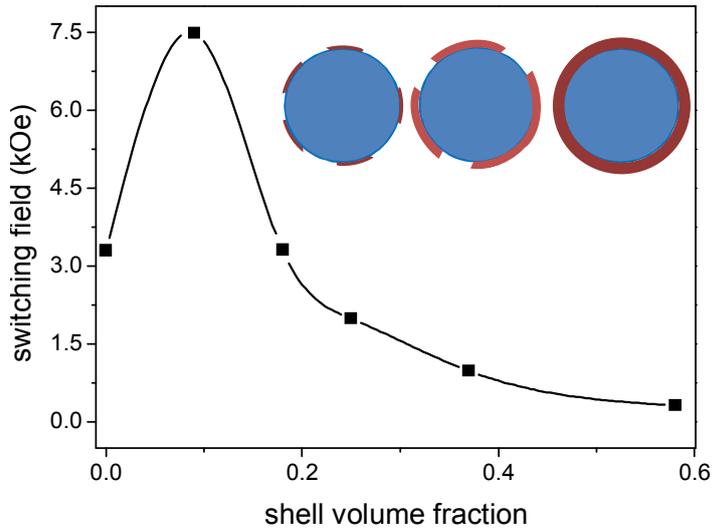

**Figure 2.** Calculated inversion of various CoFe$_2$O$_4$@MnFe$_2$O$_4$ nanoparticles, according to the volume-averaged $\frac{K_{eff}}{M_{eff}}\frac{a}{t}$ where $t$ is the thickness of the shell, $a$ the size of the unit cell. The line is a guide to the eye. Inset shows the proportion of covered core surface upon increasing the shell thickness. To allow a direct comparison with the results of ref. [3], parameters were taken from Table S3 in that paper: anisotropy constant (in units of J/m$^3$) of $K_{hard} = 2 \times 10^5$ (core) and $K_{soft} = 4 \times 10^3$ (shell); saturation magnetization (A/m) $M_{hard} = 4 \times 10^5$ (core) and $M_{soft} = 5 \times 10^5$ (shell). Though, please note the much smaller value of the effective anisotropy constant (4.82 × 10$^3$ J/m$^3$) of the nucleus given in Table S2. From our point of view the case study methodology by Moon *et al.* lacks rigor (for instance, we note eq. (1) within the Supporting Information section 3 is valid for a Stoner-Wohlfarth system with uniaxial anisotropy and only if the easy axes of all particles are aligned, whereas eq. (2) in there intrinsically assumes particles to be spherical in shape, while real samples are cuboids). For this reason coercivity at $t$ = 0 was fixed to 3.3 kOe.

## Discussion

Neither the SLP values nor the correlation of the magnetic properties with different shell volume fractions can be easily understood by examining the paper by Moon *et al.* Moreover, no correspondence of (BH)$_{max}$ with the hyperthermia potential was described. Indeed, the reported estimated heating efficiencies are, as mentioned earlier, much larger than the magnetic hysteresis loop models would predict. In this regard we note this is not the first time values of heating efficiency reported by the same group were called into question.[12, 19]

The first explanation that comes to mind is that the several issues in the paper by Moon *et al.* result from unintentional problems with experimental design or analysis. After some e-mail exchange with Prof. Cheon, the authors of the article acknowledged incongruences and a correction followed (DOI: 10.1021/acs.nanolett.7b01759). But when these errors are fixed, if the lower abscissa in Figure 4b in ref. [3] has to be scaled by 60 to get consistency, then no heat transfer model could provide an explanation to the incredible fast saturation of temperature (according to the online Supporting Information, Section 6, the solution was placed in the center of a water-cooled magnetic induction coil with a diameter of 5 cm insulated by Styrofoam). Equations describing the heat flow from the sample to the surrounding environment, of an analogous system, can be found in ref. [20]. Without the heat exchange to the surrounding medium, the temperature would increase linearly with the heating time at a specific heating rate. If the adiabatic condition is not perfectly realized, the temperature evolution is nonlinear. Eventually the system would reach the steady state. And the time constant can be estimated. Results were compared with the same group's previously reported heat transfer profiles, using the very same experimental set-up (Figure S6 from ref. [12]). And so if the re-analyzed data from the article by Moon *et al.* were to hold true, then the same group's work by Lee *et al.* on which much magnetic nanoparticles research policy has been based would, according to heat transfer standards, necessarily be impossible; and viceversa.

Last but not least, the recent correction (DOI: 10.1021/acs.nanolett.7b01759) to the work by Moon *et al.* would not be capable of solving the misinterpretation of $(BH)_{max}$. It is important to emphasize that Moon *et al.* considered the remanence in their system $M_R = 0.5M$ to be half the magnetization at saturation. We note this is true for the special case of noninteracting particles, also with the easy axes distributed at random. Which sets then a relationship between coercivity $H_C \approx 0.5 H_A$ and the anisotropy field, being $H_A = 2K/M$ in the form of uniaxial type for the sake of simplicity. But here we argue that this is not the case depicted in ref. [3] (see for instance Fig. 2b and Fig. 3b therein) where hysteresis loops are approximately square with $M_R \approx M$. In fact it resembles the case of particles with easy axes aligned along the magnetic field direction; the coercive field equals the anisotropy field, which gives a maximum hysteresis loss $A = 4 H_C M_R = 8K$ (instead of the value $\approx 2K$ for a random angular distribution of the anisotropy easy axis). This is what is expected for ordered arrays given the capacity of cuboids to self-assemble spontaneously,[21] as can be deduced from the electron microscopy images such as Fig. 1b in ref. [3]. Continuing in the spirit of order-of-magnitude calculations, let us surmise magnetic characterization was carried out in the solid powder, i.e. the precipitate after centrifugation, while the measurement of specific loss power was done after dispersing them in toluene. So, it seems feasible the dried nanoparticles "supperlattices" (what is being measured in SQUID) could eventually get transformed into elongated arrangements in solution (what is being measured during hyperthermia protocols) due to the magnetic dipole-dipole interaction. In this regard, we have previously focused on understanding the heating trends after the chain formation in the case of magnetosomes and the alike,[22] where the heating performance can be easily doubled in comparison with a randomly distributed system. Actually, hysteresis losses about A ≈ 4K would explain the factor of two between the maximum output of 180 W/g (estimated from the maximum energy product in the nanoprecipitate) and the specific loss power about 90 W/g in solution. This conclusion may receive some additional support from the fact $(BH)_{max}$ is twice the maximum magnetostatic energy available from a magnet.[23]

In summary, the claim of Moon *et al.* that a new interface phenomenon enhanced the spin canting at the core(hard)/shell(soft) interface, and which is operative when the shell is just a few atomic layers thick, is not justified. Furthermore, the large enhancement of the hyperthermia efficiency is highly questionable, and it looks like data in the publications cited above have been inappropriately manipulated. While there is some prospect of increasing the heating efficiency, the tens of kilowatts per gram ferrofluid at affordable magnetic fields seem to be just out of reach. Future work will demonstrate if this hypothesis holds true.

■ AUTHOR INFORMATION

**Corresponding Author**

*E-mail: cboubeta@gmail.com

**Notes**

The author declares that no competing interests exist.

■ ACKNOWLEDGEMENTS

C.M.B. acknowledges instructive discussions with Konstantinos Simeonidis and David Serantes.

■ REFERENCES

[1] Houlding T. K.; Rebrov, E. V. *Green Process. Synth.* **2012,** *1,* 19-31.
[2] Beck, M. M.; Lammel, C.; Gleich, B. *J. Magn. Magn. Mater.* **2017,** *427,* 195–199.
[3] Moon, S. H.; Noh, S.-h.; Lee, J.-H.; Shin, T.-H.; Lim, Y.; Cheon. J. *Nano Lett.* **2017,** *17,* 800–804.
[4] Kaptchuka, T. J.; Catherine, E. K.; Zangerb, A. *The Lancet* **2009,** *374,* 1234–1235.
[5] Toumey. C. *Nature Nanotech.* **2016,** *11,* 2–3.
[6] Meister. M. *eLife* **2016,** *5,* e17210.
[7] Baker, M. *Nature* **2016,** *533,* 452–454.
[8] Nosek, B. A.; Errington, T. M. *eLife* **2017,** *6,* e23383.
[9] Fernández van Raap, M. B.; Coral, D. F.; Yu, S.; Muñoz, G. A.; Sánchez, F. H.; Roig, A. *Phys. Chem. Chem. Phys*. **2017,** *19,* 7176-7187.
[10] Goodwin, R. D. *J. Phys. Chem. Ref. Data* **1989,** *18,* 1565-1636.
[11] Fortin J.P.; Wilhelm C.; Servais J.; Ménager C.; Bacri J. C.; Gazeau F. *J. Amer. Chem. Soc.* **2007,** *129,* 2628–2635.
[12] Lee, J.-H.; Jang, J. T.; Choi, J. S.; Moon, S. H.; Noh, S.-h.; Kim, J. W.; Kim, J. G.; Kim, I. S.; Park, K. I.; Cheon, J. *Nature Nanotechnol.* **2011,** *6,* 418-422.
[13] Nandwana, V.; Chaubey, G. S.; Yano, K.; Rong, C.; Liu, J. P. *J. Appl. Phys*. **2009,** *105,* 014303.
[14] Ali, M.; Marrows, C. H.; Hickey, B. J. *Phys. Rev. B* **2008,** *77,* 134401.
[15] Choi Y.; Jiang J. S.; Pearson J. E.; Bader S. D.; Kavich J. J.; Freeland J. W.; Liu J. P. *Appl. Phys. Lett.* **2007,** *91,* 072509.
[16] López-Ortega, A.; Estrader, M.; Salazar-Alvarez, G.; Roca, A. G.; Nogués. J. *Phys. Rep*. **2015,** *553,* 1–32.
[17] Martinez Boubeta, C.; Clavero, C.; García-Martín, J. M.; Armelles, G.; Cebollada, A.; Balcells, Ll.; Menéndez, J. L.; Peiró, F.; Cornet, A.; Toney, M. F. *Phys. Rev. B* **2005,** *71,* 014407.
[18] Ohldag, H.; Scholl, A.; Nolting, F.; Arenholz, E.; Maat, S.; Young, A. T.; Carey, M.; Stöhr, J. *Phys. Rev. Lett*. **2003,** *91,* 017203.
[19] Carrião, M. S.; Bazukis, A. F. *Nanoscale* **2016,** *8,* 8363-8377.
[20] Feng, W.; Zhou, X.; Nie, W.; Chen, L.; Qiu, K.; Zhang, Y.; He, C. *ACS Appl. Mater. Interfaces* **2015,** *7,* 4354–4367.
[21] Martinez-Boubeta, C.; Simeonidis, K.; Makridis, A.; Angelakeris, M.; Iglesias, O.; Guardia, P.; Cabot, A.; Yedra, Ll.; Estradé, S.; Peiró, F.; Saghi, Z.; Midgley, P. A.; Conde-Leborán, I.; Serantes, D.; Baldomir, D. *Sci. Rep.* **2013,** *3,* 1652.
[22] Serantes, D.; Simeonidis, K.; Angelakeris, M.; Chubykalo-Fesenko, O.; Marciello, M.; Morales, M.; Baldomir, D.; Martinez-Boubeta, C. *J. Phys. Chem. C* **2014,** *118,* 5927–5934.
[23] Skomski, R.; Coey, J. M. D. *Phys. Rev. B* **1993,** *48,* 15812-15816.